\documentclass[aps,10pt]{revtex4}

\newcommand{\be}{\begin{equation}}
\newcommand{\ee}{\end{equation}}
\newcommand{\ba}{\begin{eqnarray}}
\newcommand{\ea}{\end{eqnarray}}

\def\beq{\begin{eqnarray}}
\def\eeq{\end{eqnarray}}
\def\ln{\,\mbox{ln}\,}

\def\ga{\gamma}

\def\la{\lambda}

\def\pa{\partial}

\begin{document}
\title{Gap generation for Dirac fermions on Lobachevsky plane in a
magnetic field}


\author{E.V. Gorbar}
\email{gorbar@bitp.kiev.ua}
\author{V.P. Gusynin}
\email{vgusynin@bitp.kiev.ua}
\affiliation{Bogolyubov Institute for Theoretical Physics, 03143,
Kiev, Ukraine }


%
%
%
%
%

\begin{abstract}

We study symmetry breaking and gap generation for fermions in the 2D space of
constant negative curvature (the Lobachevsky plane) in an external covariantly
constant magnetic field in a four-fermion model. It is shown that due to the
magnetic and negative curvature catalysis phenomena the critical coupling
constant is zero and there is a symmetry breaking condensate in the chiral
limit even in free theory. We analyze solutions of the gap equation in the
cases of zero, weak, and strong magnetic fields. As a byproduct we calculate
the density of states and the Hall conductivity for noninteracting fermions that may be relevant for studies
of graphene.

\end{abstract}


\maketitle

\section{Introduction}


Dynamical symmetry breaking (DSB) and a mass (gap) generation for fermions
usually requires the presence of a strong attractive interaction \cite{Fomin}
in order to break symmetry that makes the quantative study of DSB a difficult
problem. Therefore, it is very interesting to consider the cases where DSB
takes place in the regime of weak coupling. Three such examples are known.

The first is symmetry breaking in the presence of the Fermi surface (i.e.
chemical potential of the system is nonzero).
According to the Bardeen--Cooper--Schrieffer theory of superconductivity
\cite{BCS} (or the QCD color superconductivity studies at finite baryon density
\cite{Alf}), the Fermi surface is crucial for the formation of a bound
state and a symmetry breaking condensate in the case of arbitrary small
attraction between fermions. Indeed, according to \cite{Shankar}, the
renormalization group scaling in this case is connected only with the direction
perpendicular to the Fermi surface. Therefore, the effective dimension of
spacetime is $1 + 1$ from the viewpoint of renormalization group scaling. Since
a bound state forms for arbitrary small attraction in $1 + 1$ dimension, this
implies that the critical coupling constant is zero in this case.

The second example of DSB in the regime of weak coupling is DSB in a constant
magnetic field \cite{GMSh1,GMSh3} (for a short review see \cite{UPHZH}), where
symmetry is again dynamically broken for arbitrary weak interaction. The
physical reason for this is the effective dimensional reduction of spacetime
for fermions in the infrared region by 2 units in a constant magnetic field
(DSB in a magnetic field in spacetimes of dimension higher than four was
considered in \cite{G1}). The reduction occurs because electrons being charged
particles cannot propagate in directions perpendicular to the magnetic field
when their energy is much less than the Landau gap $\sqrt{|eB|}$.

Dynamics of fermions in hyperbolic spaces $H^D$ gives the third known example
of DSB with zero critical coupling constant \cite{C1} (for an excellent review
of DSB in curved spacetime see \cite{Ina}). Analyzing the heat kernel, it was
shown in \cite{G2} that the zero value of the critical coupling constant for
DSB in hyperbolic spaces is connected with an effective dimensional reduction
in the infrared region for fermions. The combined effect of constant magnetic
field and negative curvature of spacetime on the dynamics of symmetry breaking
was studied in \cite{Odintsov}, where magnetic field was treated exactly but
gravitational field was considered in the weak curvature approximation.

In this paper, we investigate DSB and gap
generation for fermions in the $R \times H^2$ spacetime of constant negative
curvature (the Lobachevsky plane) with an external covariantly constant
magnetic field by treating gravitational and magnetic fields exactly.
The study of effects of surface curvature may be important for some condensed
matter systems \cite{Bulaev,GusyninPRL2005,Vozmediano}. In particular, the
quantum Hall effect (QHE) for two-dimensional nonrelativistic electron gas on the surface of
constant negative curvature  was studied in
Ref.\cite{Bulaev}. The present work may be relevant for the QHE in graphene
\cite{GusyninPRL2005} (see also \cite{Ando}) whose quasiparticle excitations possess the linear
dispersion law and are described by the Dirac equation (with the light velocity
replaced by the Fermi velocity). Although the Lobachevsky plane has not been
experimentally realized yet, the interaction with impurities or defects in
graphene may lead to the effective Dirac equation in a curved space. For
example, substitution of some hexagons by pentagons (heptagons) in the
hexagonal lattice leads to the warping of the graphene sheet and induces
positive (negative) curvature in it \cite{Vozmediano}. The curvature of the
graphene sheet influences the density of states and affects the transport
properties (the QHE among them) that can be observed in experiment.

In Sec.\ref{GNmodel:general} we present a general expression for the effective
potential in a four-fermion model of the Gross-Neveu type on the Lobachevsky
plane in a constant magnetic field. We then analyze  the dynamics of free
fermions and calculate the density of states and the Hall conductivity. In
particular, we show that the anomalous half-integer QHE takes place for the
Lobachevsky plane and the effect of nonzero curvature is to shift the plateau
transitions in the Hall conductivity to higher values of magnetic field and
decrease the plateau widths. The gap equation for interacting fermions is
derived in Sec.\ref{gap:eq} and its analysis is given in subsections
\ref{Zerofield}, \ref{weakfield}, \ref{strongfield}. We find solutions of the
gap equation in the cases of zero, weak, and strong magnetic fields. For the
case of zero external magnetic field, using solutions of the Dirac equation and
studying classical motion in $H^D$, we clarify in Appendix physical reasons for
the effective dimensional reduction $D + 1 \to 1 + 1$ in the infrared region
for fermions in spacetimes $R \times H^D$.


\section{Four-fermion model on the Lobachevsky plane: general analysis}
\label{GNmodel:general}

The Lobachevsky plane (or 2D hyperbolic space $H^2$) is the simplest example of
a space with constant negative curvature. Using Poincare coordinates for the
$H^2$ space, the interval on the static spacetime $R \times H^2$ is given by
\beq ds^2 = dt^2 - \frac{a^2}{y^2}(dx^2 + dy^2), \eeq where $y
> 0$ and $a$ is the curvature radius of the Lobachevsky space. The vector
potential $
\vec{A} = (\frac{Ba^2}{y}, 0)$
defines covariantly constant magnetic field on the Lobachevsky plane. Indeed,
nonzero components of the strength tensor are
\begin{eqnarray*}
F_{12}=-F_{21}=\frac{Ba^2}{y^2},
\end{eqnarray*}
and one can easily check that the corresponding $F_{\mu\nu}$ is covariantly
constant, i.e., $\nabla^{\mu} F_{\mu\nu} = 0$.

We consider a four-fermion model of the Gross-Neveu type with $N_{f}$ flavors
in $2+1$ spacetime whose action reads
\begin{equation}
     S = \int\!d^3x \sqrt{-g} \left[\sum_{k=1}^{N_{f}}
     \bar{\psi}_ki\gamma^{\mu}\nabla_{\mu}\psi_k
     + \frac{G}{2N_{f}} (\sum_{k=1}^{N_{f}}\bar{\psi}_k\psi_k )^{2} \right],
\label{H2action}
\end{equation}
where $g = \mbox{det} (g_{\mu\nu})$ is the determinant of the metric tensor,
$\nabla_{\mu} = \partial_{\mu}+ieA_\mu + i \omega^{ab}_{\mu} \sigma_{ab}$ the
covariant derivative with the spin connection $\omega^{ab}_{\mu}$, and
$\gamma^{\mu}$ matrices in curved spacetime are related to the Dirac $\gamma^a$
matrices in flat spacetime through dreibeines $\gamma^{\mu} =
e^{\mu}_{a}\gamma^{a}$, and we use a reducible, four-dimensional
representation of the Dirac algebra. Model (\ref{H2action}) (with $N_{f}=2$) in
flat space was recently proposed as the low-energy theory of interacting
electrons on graphene's two-dimensional honeycomb lattice \cite{Herbut} where a
four-fermion term arises from the microscopic lattice interactions.

The action (\ref{H2action}) is invariant with respect to $U(1)\times U(1)$
continuous transformations $\psi\rightarrow e^{i\alpha}\psi,\, \psi\rightarrow
e^{i\theta\gamma_3\gamma_5}\psi,$ and the discrete chiral transformations \be
\psi\rightarrow-i\gamma_3\psi,\quad\psi\rightarrow\gamma_5\psi.
\label{discrete_symm} \ee The mass term $m\bar\psi\psi$ would break these
discrete symmetries while keeping intact the continuous symmetries.

It is convenient to use the auxiliary field method and represent the action
(\ref{H2action}) in the equivalent form
\begin{equation}
S = \int\!d^3x \sqrt{-g} \left[ \sum_{k=1}^{N_{f}}\left(i\bar{\psi_k}
\gamma^\mu \nabla_\mu\psi_k - \sigma\bar{\psi}_k\psi_k\right) -
\frac{N_{f}}{2G}\sigma^2 \right],
\end{equation}
where $\sigma$ is an auxiliary field. If the field $\sigma(x)$ acquires a nonzero
vacuum expectation value, then fermions obtain nonzero mass and the flavor
symmetries connected with the discrete $-i\gamma_3$ and $\gamma_5$
transformations are spontaneously broken. To find the effective action for the
field $\sigma(x)$, we integrate over the fermion fields in the functional
integral. We obtain
\begin{equation}
\Gamma(\sigma) = -\int d^{3}x\sqrt{-g}\,\frac{N_{f}\sigma^{2}}{2G} -i\mbox{Ln
Det}(i\gamma^{\mu}\nabla_{\mu}-\sigma(x)).
\end{equation}
The effective potential $V(\sigma)$ is evaluated for constant field configurations ($\sigma(x)$ = const)
and is given by the expression
\begin{eqnarray*}
    V(\sigma)=-\frac{\Gamma(\sigma)}{\int d^{3}x\sqrt{-g}}.
\label{potential}
\end{eqnarray*}
Since
\begin{eqnarray}
\mbox{Det}(i\gamma^{\mu}\nabla_{\mu}-\sigma)=\mbox{Det}[\gamma_5
(i\gamma^{\mu}\nabla_{\mu}-\sigma)\gamma_5]
=\mbox{Det}(-i\gamma^{\mu}\nabla_{\mu}-\sigma),
\end{eqnarray}
we find
\begin{eqnarray*}
\mbox{Ln Det}(i\gamma^{\mu}\nabla_{\mu}-\sigma) = \frac{1}{2}
\mbox{Tr Ln} (\partial_0^2 + D^2 +\sigma^2)=
-\frac{1}{2}\int\limits_0^\infty\frac{dt}{t}\,{\rm
Tr}\exp[{-i(\partial_0^2 + D^2 +\sigma^2)t}],
\end{eqnarray*}
where $D=\vec{\gamma}\vec{\nabla}$, the trace ${\rm Tr}$ is taken in the
functional sense and we used the formal identity
$\ln(H-i\epsilon)=-\int\limits_0^\infty\exp[-it(H-i\epsilon)]{dt}/{t}$ for the
logarithm of an operator $H$. Further we calculate \ba \mbox{Ln
Det}(i\gamma^{\mu}\nabla_{\mu}-\sigma)=-\frac{1}{2}\int\limits_{-\infty}^\infty
\frac{dk_0}{2\pi}\int\limits_0^\infty\frac{dt}{t}\,e^{ik_0^2t} {\rm
Tr}e^{-i(D^2 +\sigma^2)t} =\frac{-iN_{f}}{2(4\pi)^{1/2}}\int d^3x\sqrt{-g}
\int\limits_0^{\infty}\frac{ds}{s^{3/2}} {\mbox{tr}
\langle\mathbf{x}|e^{-s(D^2+\sigma^2)}|\mathbf{x}\rangle} , \label{log} \ea
where the trace ${\rm tr}$ goes over the Dirac indices and in the second
equality we deformed the integration contour $t\to-is$.

The gap equation $dV/d\sigma =0$ determines  a dynamically generated mass of fermions
but before analyzing it we consider the dynamics of free fermions on the Lobachevsky plane
in the presence of an external magnetic field.

\subsection{Free fermions on the Lobachevsky plane}
\label{freefermions}

The energy spectrum and eigenfunctions of the Dirac operator on the Lobachevsky
plane in an external covariantly constant magnetic field were found by Comtet
and Houston \cite{Comtet}. Comparing with the Landau problem in flat space, the
nonzero curvature of the Lobachevsky plane qualitatively changes the energy
spectrum which consists of the discrete part (we assume $eB > 0$)
\begin{equation}
E_n=\pm \sqrt{\sigma^2+\frac{b^2-(n-b)^2}{a^2}},
\end{equation}
where $n=0,1,...$, $0 \le n<b\,$, $b=eBa^2$, and the continuum part
\begin{equation}
E_{\nu}=\pm \sqrt{\sigma^2+\frac{b^2+\nu^2}{a^2}},
\end{equation}
where $0 \le \nu < \infty$. In the limit $a \to \infty$, the continuum part of
the spectrum disappears (goes to infinity) and $E_n \to \pm \sqrt{\sigma^2 +
2neB}$, i.e. the spectrum coincides with the spectrum of the Landau problem in
flat space.

The eigenfunctions of the discrete spectrum are
\begin{eqnarray}
&&\psi^{(d)}_{0,k}(x,y)=\frac{e_{-}}{a}f_0(k,b+\frac{1}{2};x,y),\quad k <0,\nonumber\\
&&\psi^{(d)}_{\alpha,n,k}(x,y)=\frac{e_{\alpha}}{a}f_{n+\frac{\alpha-1}{2}}(k,b -
\frac{\alpha}{2};x,y),\quad
k < 0, \,\, n=1,2,..., \,\, 1 \le n < b.
\end{eqnarray}
Here $\alpha$ takes values +1 and -1 and $\{e_{+},e_{-}\}$ consists of
orthonormal constant eigenvectors of $\gamma^1\gamma^2$:
$
\gamma^1\gamma^2 e_{\pm} = \pm ie_{\pm},
$
and the functions $f_n$ are
\begin{eqnarray}
f_n(k,\beta;x,y)=\sqrt{\frac{n!(2\beta-2n-1)}{4\pi|k|\Gamma(2\beta-n)}}
 e^{-ikx} e^{-|k|y}(2|k|y)^{\beta-n} L_n^{(2\beta-2n-1)}(2|k|y),
\end{eqnarray}
where $L_n^{(\alpha)}(z)$ denote associated Laguerre polynomials \cite{Grad}.
Note that for fixed $k$, the lowest level of the discrete spectrum is not
degenerate unlike the higher levels with $n \ge 1$ which are twice degenerate, similar to
the case of spectrum of the Landau problem in flat space.

The eigenfunctions of the continuum spectrum are
\begin{equation}
\psi^{(c)}_{\alpha,\nu,k}(x,y)=\frac{e_{\alpha}}{a}f(\nu,k,|b-\frac{\alpha}{2}|;x,y),
\quad k(b-\frac{\alpha}{2}) > 0,
\end{equation}
where
\begin{eqnarray}
f(\nu,k,\beta;x,y) &=& \sqrt{\frac{\nu\sinh(2\pi \nu)}{4\pi^3|k|}}
|\Gamma(i\nu-\beta+\frac{1}{2})| e^{-ikx}
z^{1/2}e^{-z/2}
\left[\frac{\Gamma(i\nu)}{\Gamma(\frac{1}{2}+i\nu-\beta)}z^{-i\nu}M(\frac{1}{2}-i\nu-\beta,
1-2i\nu;z)\right.\nonumber\\
&+&\left.\frac{\Gamma(-i\nu)}{\Gamma(\frac{1}{2}-i\nu-\beta)}z^{i\nu}M(\frac{1}{2}+i\nu-\beta,
1+2i\nu;z)\right],\quad z=2|k|y,
\end{eqnarray}
and $M(a,b;z)$ is the confluent hypergeometric function. Using the above eigenfunctions
the heat kernel of the operator $D^2+\sigma^2$ in (\ref{log}) is calculated to be (see
Ref.\cite{Comtet}),
\ba
&&\mbox{tr} <\mathbf{x}|e^{-s(D^2+\sigma^2)}|\mathbf{x}>
= \frac{1}{\pi a^2}
\left\{ be^{-s\sigma^2} + 2\sum_{n=1}^{[b]}
(b-n)e^{-s(\sigma^2 + \frac{2nb-n^2}{a^2})}\right.\nonumber\\
&&+\left. \frac{1}{\pi} \int_0^{\infty} {d\nu \nu \, e^{-s(\sigma^2 +
\frac{b^2+\nu^2}{a^2})}}\mbox{Im} \left[ \psi(i\nu-b)+\psi(i\nu-b+1)
 + \psi(i\nu+b+1) + \psi(i\nu+b) \right] \right\},
\label{heatkernel}
\ea
where $[b]$ is the largest integer satisfying $[b]\le b$
(note that our expression differs from that in \cite{Comtet} by factor 2
because we consider a reducible representation of the Dirac algebra). The
series representation for the imaginary part of $\psi$-function \cite{Grad} yields
\ba
&&\mbox{Im} \left[ \psi(i\nu-b)+\psi(i\nu-b+1)
 + \psi(i\nu+b+1)+ \psi(i\nu+b) \right]= \sum_{k=-\infty}^{+\infty}
\frac{2\nu}{\nu^2+(k-b)^2}. \label{sum-of-psi}
\ea
and the sum in Eq.(\ref{sum-of-psi}) is evaluated to be
\begin{equation}
\sum_{k=-\infty}^{+\infty}\frac{2\nu}{\nu^2+(k-b)^2} =  \frac{2\pi\sinh(2\pi
\nu)} {\cosh(2\pi \nu) - \cos(2\pi b)}. \label{sum}
\end{equation}
Finally, we obtain
\ba
\mbox{tr} \langle\mathbf{x}|e^{-s(D^2+\sigma^2)}|\mathbf{x}\rangle =\frac{1}{\pi a^2}
\hspace{-1mm}\left[ be^{-s\sigma^2}\hspace{-2mm}+ 2\sum_{n=1}^{[b]}
(b-n)e^{-s(\sigma^2 + \frac{2nb-n^2}{a^2})}
+ 2\hspace{-1mm} \int\limits_0^{\infty}\hspace{-1mm} {d\nu \nu \, e^{-s(\sigma^2 +
\frac{b^2+\nu^2}{a^2})}}
 \frac{\sinh(2\pi \nu)}
{\cosh(2\pi\nu) - \cos(2\pi b)} \right].
\label{heatkernel-final}
\ea

\subsection{Density of states and Hall conductivity}
\label{DOS-and-Hall}

Using the fermion Green's function
$$
G(\mathbf{x},\mathbf{x^{\prime}};E+i\epsilon)=\langle \mathbf{x}|
\frac{1}{\gamma^{0}(E+i\epsilon) -i{\vec \gamma}{\vec
D}-\sigma}|\mathbf{x^{\prime}} \rangle\,,
$$
we calculate the density of states (DOS) for noninteracting theory which is
defined as \ba \rho(E)=-\frac{N_{f}}{\pi V}\int d^{2}x{\rm tr}[\gamma^{0}{\rm
Im}G(\mathbf{x},\mathbf{x};E+i\epsilon) ]=\frac{N_{f}}{\pi}{\rm Im}\,{\rm
tr}\langle \mathbf{x}|\frac{E}{D^{2}+\sigma^{2}-(E+i\epsilon)^{2}} |\mathbf{x}
\rangle, \ea where $V$ is space volume which is canceled because diagonal
matrix elements of the operator $D^2$ do not depend on $\mathbf{x}$ (see
Eq.(\ref{heatkernel-final})) in view of homogeneity of the Lobachevsky space.
In order to calculate the DOS we integrate Eq.(\ref{heatkernel-final}) over $s$
from $0$ to $\infty$, replace $\sigma^{2}$ by $\sigma^{2}-(E+i\epsilon)^{2}$,
and take the imaginary part. We get \ba
&&\rho(E)=\frac{N_{f}eB}{2\pi}[\delta(E-\sigma)+\delta(E+\sigma)]
+\frac{N_{f}}{\pi a^{2}}\sum\limits_{n=1}^{[b]}(b-n)
[\delta(E-E_{n})+\delta(E+E_{n})]\nonumber\\
&&+\frac{N_{f}|E|}{\pi}\,\theta\left(|E|-
\sqrt{\sigma^{2}+\frac{b^{2}}{a^{2}}}\right)\frac{\sinh(2\pi\nu(E))}
{\cosh(2\pi\nu(E))-\cos(2\pi b)},
\label{DOS}
\ea
where
\ba
&&E_{n}=\sqrt{\sigma^{2}+\frac{2nb-n^{2}}{a^{2}}},\quad
\nu(E)=\sqrt{a^{2}(E^{2}-\sigma^{2})-b^{2}}.
\ea
The first two terms in Eq.(\ref{DOS}) correspond to the point spectrum and the last term
corresponds to the continuous one.

One can easily check that in flat space ($a=\infty$) the DOS given by
Eq.(\ref{DOS}) reduces to the corresponding expression (4.2) in
Ref.\cite{oscillations}. On the other hand, in the absence of a magnetic field
($B=0$) we have
\be
\rho(E)=\frac{N_{f}|E|}{\pi}\,\theta(|E|-\sigma)\coth(\pi
a\sqrt{E^{2}-\sigma^{2}}). \label{DOS-0}
\ee
For massless fermions ($\sigma \to
0$) we find that the DOS (\ref{DOS-0}) remains finite at $E=0$,
\be
\rho(0)=\frac{N_{f}}{\pi^{2}a},
\label{zerofield-DOS}
\ee
in contrast to the flat space where it
vanishes like $\rho(E)\sim |E|$ when $E\to0$. It also differs from
nonrelativistic fermions on the Lobachevsky plane where the DOS behaves as
$\rho(E)\sim \sqrt{|E|}$ when $E\to0$ \cite{Bulaev}.

In the case of high magnetic fields and large radius of curvature
($e^2B^{2}a^{2}>\mu^{2}-\sigma^{2}$) the energy spectrum below the Fermi level
$\mu$ is only the discrete one, therefore the last term in Eq.(\ref{DOS})
related to the continuum spectrum does not contribute. In this case the number
of states below the Fermi level $N(\mu)$ can be easily calculated using
Eq.(5.23) at zero temperature in Ref.\cite{oscillations},
\ba
N(\mu)=V{\rm
sign}(\mu)\int\limits_{0}^{|\mu|}dE\rho(E)=\frac{N_{f}V}{2\pi a^{2}}{\rm
sign}(\mu)\left(b+2(b-\frac{n_{max}+1}{2})n_{max}\right),
\ea
where $n_{max}$ equals the maximal filled state number and is given by the integer part of the
following expression:
\begin{eqnarray*}
n_{max}=[b-\sqrt{b^{2}-a^{2}(\mu^{2}-\sigma^{2})}].
\end{eqnarray*}
Using $N(\mu)$, we can calculate the Hall conductivity through the
Streda formula \cite{Streda},
\be
\sigma_{xy}(\mu,B)=-\frac{e}{V}\frac{\partial N(\mu)}{\partial B},\,
e>0,
\ee
which is valid when the Fermi level lies within an energy
gap. In the energy gap the integer part of $b-\sqrt{b^{2}-a^{2}(\mu^{2}-\sigma^{2})}$ is constant
and we obtain
\ba \sigma_{xy}(\mu,B)=-{\rm
sign}(\mu)\frac{N_{f}e^{2}}{\pi} \left(\frac{1}{2}+
[b-\sqrt{b^{2}-a^{2}(\mu^{2}-\sigma^{2})}]\right).
\ea
This expression describes the Hall conductivity of relativistic fermions
on the Lobachevsky plane. As is seen, the field dependence of the
Hall conductivity has a familiar step-like behavior. In the limit of
zero curvature ($a\to\infty$) we get \ba \sigma_{xy}(\mu,B)=-{\rm
sign}(\mu)\frac{N_{f}e^{2}}{\pi}\left(\frac{1}{2}+[\frac{\mu^{2}
-\sigma^{2}}{2eB}]\right). \ea For massless fermions, $\sigma=0$,
this expression coincides (for $N_{f}=2$ and restoring the constants
$\hbar, c$ and the Fermi velocity $v_{F}$) with Eq.(7) in
Ref.\cite{GusyninPRL2005} which describes the unconventional quantum
Hall effect in graphene. The anomalous QHE with half-integer
quantization of the Hall conductivity for Dirac fermions is due to
the lowest  Landau level whose degeneracy is two times less than
degeneracy of any other level. Notice that this anomalous QHE is the
most direct evidence for the existence of Dirac fermions in graphene
\cite{Novoselov}. The half-integer quantization of the Hall
conductivity remains valid for the Lobachevsky plane (for the QHE
exhibited by relativistic particles on a two-sphere see recent paper
\cite{Jellal}). The effect of nonzero negative curvature is to shift
the plateau transitions (which arise from the crossings of the Fermi
level with the Landau levels) in the Hall conductivity to higher
magnetic fields,
\be
B_{n}=\frac{\mu^{2}-\sigma^{2}}{2en}+\frac{n}{2ea^{2}}, \ee and
decrease the plateau widths, \be \Delta
B=\frac{\mu^{2}-\sigma^{2}}{2en(n+1)}-\frac{1}{2ea^{2}},
\ee
similar to the case of nonrelativistic electrons \cite{Bulaev}.

\section{Gap equation}
\label{gap:eq}
We now turn to the analysis of the gap equation ${dV}/{d\sigma} = 0$ which takes the following
form:
\begin{equation}
\sigma =\frac{G}{\pi a^2} \left[ \frac{b}{2} + \sum_{n=1}^{[b]}
\frac{b-n}{\sqrt{1 + \frac{2nb-n^2}{\sigma^2a^2}}}
 + \int_0^{\Lambda a} \frac{d\nu \nu}{\sqrt{1 +
\frac{b^2+\nu^2}{\sigma^2a^2}}} \frac{\sinh(2\pi \nu)}{\cosh(2\pi \nu) -
\cos(2\pi b)} \right], \label{gapequation}
\end{equation}
where $\Lambda$ is the ultraviolet (UV) cut-off.
In what follows we find approximate analytical solutions of the gap equation
(\ref{gapequation}) in the cases $b=0$, $b \ll 1$, and $b \gg 1$.

It is instructive before solving the gap equation to calculate the symmetry
breaking condensate $<0|\bar{\psi}\psi|0>$ in the chiral limit $\sigma\to 0$.
Naively one would expect that it is zero in the chiral limit. However, like in
the case of free fermions in flat $(2+1)$-dimensional spacetime with $B \ne 0$
\cite{GMSh1} the condensate is nonzero in the chiral limit in the case under
consideration. The reason is that, although the Landau spectrum is modified for
fermions in a covariantly constant magnetic field in $H^2$ (see \cite{Comtet}),
the lowest zero level still survives and this leads to the effective reduction
of spacetime dimension by two units and, as result, to a nonzero condensate.
The condensate in our model equals \ba \langle 0|\bar{\psi}\psi|0\rangle = -
\lim_{x\to x^\prime}\mbox{tr}\, G(x,x^\prime) =
-\frac{N_{f}{\sigma}}{2\sqrt{\pi}} \int\limits_0^{\infty} \frac{ds}{\sqrt{s}}
\,\mbox{tr} \langle{\vec x}|e^{-s(D^2+{\sigma}^2)}|{\vec x}\rangle, \ea where
$G(x,x^\prime)$ is the fermion Green's function. Using the heat kernel
Eq.(\ref{heatkernel}), we find that only the lowest Landau level contributes to
the condensate in the chiral limit ${\sigma}\to0$:
\begin{equation}
<0|\bar{\psi}\psi|0> = -N_{f}{\sigma} \int_0^{\infty} \frac{ds}{\sqrt{4\pi s}}
\frac{be^{-s{\sigma}^2}}{\pi a^2}= -\frac{N_{f}eB}{2\pi}. \label{condensate}
\end{equation}
Note that condensate (\ref{condensate}) does not depend on curvature of
spacetime and exactly coincides with the corresponding flat spacetime result
\cite{GMSh1}.

\subsection{Zero Magnetic Field}
\label{Zerofield}

In this section we consider solutions of the gap equation for zero magnetic
field making an accent on the physics underlying the phenomenon of spontaneous
mass generation in hyperbolic spaces. Unlike the dimensional regularization
usually considered in the literature, we use the regularization with an
explicit UV cut-off.

For $B=0$, the gap equation (\ref{gapequation}) significantly simplifies and we get
\begin{equation}
\sigma  = \frac{G\sigma}{\pi a} \int_0^{\Lambda a} \frac{d\nu \nu}{\sqrt{{\sigma^2a^2}
+ {\nu^2}}} \coth(\pi \nu),
\label{ge0}
\end{equation}
which up to the terms of order $1/\Lambda$ can be rewritten in the form
\begin{equation}
\sigma =\frac{G\sigma}{\pi}\left(\Lambda-\sigma \right)+\frac{G\sigma }{\pi^2 a}\int_0^\infty
\frac{d\nu\nu(\coth\nu-1)}{\sqrt{\nu^2+(\pi \sigma a)^2}}.
\label{gap_zero_field}
\end{equation}
As is seen, there always exists the trivial solution $\bar\sigma =0$. In the
case of flat space ($a=\infty$) the integral on the right hand side of
Eq.(\ref{gap_zero_field}) vanishes and the equation admits a nontrivial
solution only if the coupling constant $G$ is supercritical,
$G>G_c=\pi/\Lambda$. However, on the Lobachevsky plane with the finite
curvature radius $a$ the situation changes dramatically: a nontrivial solution
exists for all $G>0$. The reason for this is that in the Lobachevsky space the
interaction becomes enhanced in the infrared region (small $\nu$): the integral
in Eq.(\ref{gap_zero_field}) is proportional to $\ln(1/(\pi \sigma a))$ as
$\sigma \to0$. Analytical solution can be obtained for the coupling constant
$G<<a$. We find \be m_{dyn}\equiv \bar\sigma =\frac{1}{\pi
a}\exp\left(-\frac{\pi^2a}{G}\right). \label{mass_dyn_zerofield} \ee Note that
this solution does not depend on the ultraviolet cutoff $\Lambda$. This
confirms that the effect of mass generation in the Lobachevsky space in the
weak coupling regime is of purely infrared origin.

To study the case of a strong coupling we introduce the dimensionless coupling
$G=\pi g/\Lambda$ and the scale $m^*= \Lambda(1/g_c - 1/g)$ where we defined
the critical coupling in the flat space $g_c=1$. Then Eq.(\ref{gap_zero_field})
is rewritten in the form \be \pi m^*a=\pi\sigma a-\int_0^\infty
\frac{d\nu\nu(\coth\nu-1)}{\sqrt{\nu^2+(\pi \sigma a)^2}}. \label{gapeq_flat}
\ee In the near critical region $g-g_c<<1/\Lambda a$ the scale $m^*\simeq0$ and
the dynamical mass is given by the root of the right hand side of
Eq.(\ref{gapeq_flat}), \be m_{dyn}=\bar\sigma=\frac{0.8}{\pi a}.
\label{scaling_law} \ee Thus in the scaling region $g-g_c<<1/\Lambda a$, the
cutoff disappears from the observable quantity $m_{dyn}$. The critical value
$g_c=1$ is in fact an UV stable fixed point at the leading order in $1/N_{f}$
expansion and the relation (\ref{scaling_law}) can be considered as a scaling
law in the scaling region.

In the supercritical region $g>g_c$, the analytical expression for $m_{dyn}$
can be obtained at large curvature radius $a$, satisfying the condition
$m^*a>>1$ (note that $m^*\,$ is the solution of the gap equation
(\ref{gapeq_flat}) for $a \to \infty$). Expanding the integral on the right
hand side of Eq.(\ref{gapeq_flat}) in $1/\pi \sigma a$ we find \be
m_{dyn}=\bar\sigma=m^*\left(1+ \frac{1}{12(m^*a)^2}\right),
\label{correction-flatspace} \ee i.e., $m_{dyn}$ increases with the decrease of
the curvature radius $a$. In fact a numerical study of Eq.(\ref{gapeq_flat})
shows that the dynamical mass $m_{dyn}$ increases with the decrease of $a$ for
all values of $g$ and $a$.

It is instructive to compare solution (\ref{mass_dyn_zerofield}) with the
relation for the dynamical mass in the ($1+1$)-dimensional Gross-Neveu (GN)
model \cite{Gross_Neveu} and with the quasiparticle gap in the BCS theory of
superconductivity \cite{BCS}. The relation for the dynamical mass in the
Gross-Neveu model is \be
m_{dyn}=\Lambda\exp\left(-\frac{\pi}{N_{f}G^{(0)}}\right), \label{dynmass_GN}
\ee where $G^{(0)}$ is the bare coupling, which is dimensionless for $D=1+1$.
The similarity between Eqs.(\ref{mass_dyn_zerofield}) and (\ref{dynmass_GN}) is
evident: $1/\pi a$ and $G/\pi a$ in Eq.(\ref{mass_dyn_zerofield}) play the role
of an ultraviolet cutoff and the dimensionless coupling constant in
Eq.(\ref{dynmass_GN}), respectively. This reflects the point that the dynamics
of fermion pairing in the Lobachevsky space is essentially ($1+1$)-dimensional.

We recall that in the theory of superconductivity due to the presence of the
Fermi surface the dynamics of electrons is also effectively
($1+1$)-dimensional. The analogy with the superconductivity theory is even
deeper than with the GN model. Indeed, the energy gap in the BCS theory has the
form $\Delta\sim\omega_D\exp\left(-const/\nu_SG_S\right)$, where $\omega_D$ is
the Debye frequency, $G_S$ is a coupling constant and $\nu_F$ is the density of
states on the Fermi surface. In the present model  $\rho(E=0)=N_{f}/\pi^2 a=\nu_0$,
where $\nu_0$ is the density of states on the energy surface $E=0$
(see Eq.(\ref{zerofield-DOS})). Thus the energy surface
$E=0$ plays here the role of the Fermi surface. Hence
Eq.(\ref{mass_dyn_zerofield}) can be rewritten in the form $m_{dyn}=({1}/{\pi
a})\exp\left(-{N_{f} }/{\nu_0G}\right)$ exhibiting a complete analogy with the
energy gap in the BCS theory. Moreover, it can be shown that the effective
dimensional reduction $D + 1 \to 1 + 1$ for fermions in the infrared region
takes place for $R\times H^D$ spacetimes of any dimension $D\ge2$ and the same
reduction $D+1\to1+1$ is valid for the BCS theory in $D+1$ dimension. In
Appendix we present physical reasons for the reduction $D+1\to1+1$ in
hyperbolic spacetimes $R\times H^D$ studying the classical motion of particles
and solving also the Dirac equation in these spacetimes.

\subsection{Weak Magnetic Field}
\label{weakfield}
To study the dynamical mass generation in the Lobachevsky space in the presence
of an external magnetic field we first rewrite Eq.(\ref{gapequation}) in more
convenient form,
\begin{eqnarray}
\sigma=\frac{G\sigma}{\pi}(\Lambda-\sigma)+\frac{G\sigma}{\pi a}\left[\frac{b}{2\sigma a}+
\sum\limits_{n=1}^{[b]}\frac{b-n}{\sqrt{(\sigma a)^2+2bn-n^2}}\right.
+\left.\int\limits_0^\infty\frac{d\nu\nu}{\sqrt{\nu^2+b^2+(\sigma
a)^2}} \left(\frac{\sinh(2\pi \nu)}{\cosh(2\pi \nu) - \cos(2\pi
b)}-1\right)\right],
\label{gapeq_magfield}
\end{eqnarray}
up to terms of order $1/\Lambda$. For a weak external magnetic field when $b
\ll {\rm min}\,(1,\,\sigma a)$ or equivalently the magnetic length
$l=1/\sqrt{eB}$ satisfying $l>>{\rm max}\,(a,\,\sqrt{\frac{a}{\sigma}})$, we
neglect $b^2$ and higher order terms in Eq.(\ref{gapeq_magfield}) and obtain
the following gap equation:
\be
\pi\sigma a=\pi m^*a+\frac{\pi b}{2\sigma
a}+\int\limits_0^\infty \frac{d\nu\nu(\coth\nu-1)}{\sqrt{\nu^2+(\pi \sigma
a)^2}}, \label{ge10}
\ee
where the second term on the right hand side
represents, obviously, a first order correction to the gap equation
(\ref{gapeq_flat}). Seeking the solution in the form $\sigma = m_{dyn}^{(0)} +
Cb$ where $m_{dyn}^{(0)}$ is the solution of the zero field gap equation
(\ref{gapeq_flat}), we find
\be
m_{dyn}\equiv\bar\sigma=m_{dyn}^{(0)}\left(1+\frac{eB}{2(m_{dyn}^{(0)})^2}\right),
\ee
i.e., the dynamical mass always increases with $B$.
A striking fact is that,
unlike the gap equation (\ref{gap_zero_field}) with $B=0$, the gap equation
(\ref{gapeq_magfield}) with $B\neq0$ does not have the trivial solution
$\bar{\sigma}=0$. Thus, despite the spontaneous character of breaking of the
discrete symmetries (\ref{discrete_symm}), there is no trivial solution in the
magnetic field for all values of the coupling constant $G$, the fact already
known in the case of a flat space \cite{GMSh3}.

\subsection{Strong Magnetic Field}
\label{strongfield}
For strong magnetic field $b \gg 1$, we can determine the leading asymptotics
of the sum in (\ref{gapeq_magfield}) as $b \to \infty$. We have
\begin{equation}
\sum_{n=1}^{b} \frac{b-n}{\sqrt{\sigma^2a^2 + {2nb-n^2}}} =
b\sum_{n=1}^{b} \frac{1}{b} \frac{1-\frac{n}{b}}{\sqrt{\frac{\sigma^2a^2}{b^2} +{\frac{2n}{b}-\frac{n^2}{b^2}}
}}.
\label{sum1}
\end{equation}
For $b \to \infty$, we can make the change $\frac{1}{b} \to dx$ and replace the
sum over $n$ by integral over $x$. Then sum (\ref{sum1}) is approximated by the
integral
\begin{eqnarray*}
I = b\int_{\frac{1}{b}}^1 dx \frac{1-x}{\sqrt{\frac{\sigma^2a^2}{b^2} + 2x-x^2}}
 = \sqrt{b^{2}+\sigma^2a^2}-\sqrt{2b-1+\sigma^2a^2}.
\label{sum2}
\end{eqnarray*}
Consequently, we obtain the following gap equation:
\begin{eqnarray}
\sigma = \frac{G}{\pi a^2} \left[ \frac{b}{2} + \sigma aI +\Lambda\sigma a^2 - \sigma a
\sqrt{b^2+(\sigma a)^{2}}+\sigma a
\int\limits_{0}^{\infty}\frac{d\nu\nu}{\sqrt{\nu^{2}
+b^{2}+(\sigma a)^{2}}}
\left(\frac{\sinh(2\pi\nu)}{\cosh(2\pi\nu)-\cos(2\pi b)}-1\right) \right].
\label{ge21}
\end{eqnarray}
Let us now consider solutions of this gap equation for $\sigma a \ll b$ and
$\sigma a \gg b$. For $\sigma a \ll b$, $I \simeq b-\sqrt{2b}$ and we can
neglect the integral in Eq.(\ref{ge21}) which is of order $1/b$. Thus, we find
the solution
\begin{equation}
m_{dyn}=\bar{\sigma} = \frac{{Gb}/{(2\pi a^2)}}{1 -
\frac{G\Lambda}{\pi}+\frac{G\sqrt{2eB}}{\pi }} \approx \frac{{Gb}/{(2\pi
a^2)}}{1 - \frac{G\Lambda}{\pi}}. \label{wcs}
\end{equation}
This solution is obviously valid for
$
G < {\pi}/{\Lambda},
$
i.e., this solution corresponds to the weak coupling regime. One can check that the condition
$\sigma a \ll b$ is also satisfied because
\begin{equation}
m_{dyn}=\bar{\sigma} \approx \frac{Gb}{2\pi a^2} = \frac{GeB}{2\pi}
\label{flatspacesolution}
\end{equation}
in the weak coupling regime. Solution (\ref{flatspacesolution}) is exactly the
flat spacetime solution in $2+1$ dimension in a constant external magnetic
field \cite{GMSh1}. Of course, this is a natural result because for strong
magnetic field $eB \gg {1}/{a^2}$, we can neglect small corrections due to
the curvature of space. Further, one can check that solution
(\ref{flatspacesolution}) satisfies the condition $\bar{\sigma} a \ll b$
because  $G \ll 2\pi a$.

For the other case $\sigma a \gg b$, $I \simeq b^{2}/(2 \sigma a)$ and we find
\begin{equation}
\sigma \approx \frac{G}{\pi a^2}(\frac{b^2}{2} + \sigma\Lambda a^2 - \sigma^2a^2),
\label{scs1}
\end{equation}
that gives
\begin{equation}
m_{dyn}=\bar{\sigma} = \frac{m^{*}+ \sqrt{{m^{*}}^2+\frac{2b^2}{a^2}}}{2}.
\label{scs2}
\end{equation}
The condition $\sigma a \gg b$ implies that ${m^{*}}^2=(\Lambda - \frac{\pi}{G})^2$ should be much larger than
${2b^2}/{a^2}$. Therefore, this solution exists for $G > {\pi}/{\Lambda}$, i.e. it is
a strong coupling solution. Using the strong coupling solution
$m^{*} = \Lambda - \frac{\pi}{G}$ in flat space ($a=\infty$), we can represent
Eq.(\ref{scs2}) as follows:
\begin{equation}
m_{dyn}\equiv \bar{\sigma} \approx m^{*} + \frac{b^2}{2{m^{*}}^{2}a^2} = m^{*} + \frac{e^2B^2a^2}{2m^{*}}.
\label{scs3}
\end{equation}
Thus, we conclude that for strong external magnetic field $b \gg 1$ the
dynamical fermion mass coincides with the $2+1$-dimensional flat spacetime
solution (\ref{flatspacesolution}) in the weak coupling regime and, according
to Eq.(\ref{scs3}), the correction to the strong coupling solution $m^{*}$ due
to external magnetic field is quadratic in $B$.

\section{Conclusion}
\label{concl} In the present paper we studied a four-fermion Gross-Neveu type
model on the Lobachevsky plane in an external covariantly constant magnetic
field. For noninteracting fermions we calculated the density of states and the Hall
conductivity which may be relevant for experimental investigations of graphene.
In particular, we showed that the density of states for free massless fermions
is finite at zero energy in contrast to the case of flat space where it
vanishes. The anomalous half-integer quantization of the Hall conductivity
remains valid for the Lobachevsky plane and the effect of nonzero negative
curvature is to shift the plateau transitions  in the Hall conductivity to
higher magnetic fields and decrease the plateau widths.

Studying the dynamical symmetry breaking we showed that discrete symmetries of
this model are always spontaneously broken, i.e. the critical coupling constant
is zero. Moreover, we found that there is a symmetry breaking condensate even
in noninteracting free theory in the chiral limit. These facts are consequences
of the effective dimensional reduction for fermions in the infrared region to a
$(0 + 1)$-dimensional theory. It is interesting that the condensate does not
depend on the value of curvature of the Lobachevsky plane and exactly coincides
with its value in flat $(2 + 1)$-dimensional space in an external constant
magnetic field \cite{GMSh1}. We analyzed the gap equation and found its
solutions for the cases of zero, weak, and strong magnetic fields.

For zero magnetic field, we showed that the dynamical mass essentially depends
on the radius of curvature of $H^2$ in the weak coupling regime $(G \ll
{\pi}/{\Lambda})$. In the strong coupling regime $(G > {\pi}/{\Lambda})$, the
dynamical mass is virtually independent of the curvature of space and
practically coincides with the flat space solution up to corrections of order
$1/(m^{*}a)^{2}$ (Eq.(\ref{correction-flatspace})). For weak magnetic field
$(eB \ll {1}/{a^2})$, the correction to the dynamical mass due to external
magnetic field is linear in $B$ and the dynamical mass grows with magnetic
field more quickly in the weak coupling regime than in the strong coupling
regime. For strong magnetic field $(eB \gg {1}/{a^2})$, we found that up to
negligible corrections, the dynamical mass coincides with the dynamical mass in
flat spacetime in an external constant magnetic field in the weak coupling
regime. In the strong coupling regime,  the correction to the flat spacetime
solution due to external magnetic field is quadratic in $B$.

It was shown in \cite{G2} that the zero value of the critical coupling constant
for DSB in $R \times H^D$ spacetimes is due to the effective dimensional
reduction $D + 1 \to 1 + 1$ for fermions in the infrared region which takes
place for any $D \ge 2$. In order to clarify physical reasons for this
reduction, we considered in Appendix solutions of the Dirac equation on $R
\times H^D$, where we showed that due to the spherical and scale symmetries the
initial Dirac problem is reduced to an effective $(1 + 1)$-dimensional problem.
Further, according to \cite{APNY}, if dimensional reduction in the infrared
region is observed in a quantum problem, then classical motion should have a
bounded character with respect to the coordinates over which the reduction
takes place, i.e. the physical system should effectively be of a finite size
with respect to these coordinates. Studying the classical motion on $H^D$, we
showed in Appendix that this is indeed the case.

\section*{Acknowledgments}

The authors are grateful to V.A. Miransky and S.G. Sharapov for useful remarks
and suggestions and acknowledge helpful discussions with S.D. Odintsov. This
work was supported by the SCOPES-project IB 7320-110848 of the Swiss NSF and
partially by the grant 10/07-H "Nanostructure systems, nanomaterials,
nanotechnologies" and by the Program for Fundamental Research of the Physics
and Astronomy Division of the National Academy of Sciences of Ukraine.

\appendix
\section{}



The metric of static $R \times H^D$ spacetime is given by \beq ds^2=dt^2 -
\frac{a^2}{x_1^2}(dx_1^2 + dx_2^2 + ... + dx_D^2),\,\,\, x_1 > 0.
\label{HDmetric} \eeq The Dirac equation in this spacetime when magnetic field
is absent has the form
\begin{equation}
\left(i\ga^0\pa_0 + \frac{ix_1}{a}\ga^1\pa_1 + ... + \frac{ix_1}{a}\ga^D\pa_D
 - \frac{i(D-1)}{2a}\ga^1 - m\right)\psi = 0.
\label{HDDequation}
\end{equation}
As usual, it is more convenient to work with the second order differential equation,
which is obtained multiplying
(\ref{HDDequation}) by $i\hat{D} + m$
\begin{equation}
(-\pa_0^2 + \frac{(D-1)^2}{4a^2} - \frac{D-2}{a^2} x_1\pa_1
+ \frac{x_1^2}{a^2}(\pa_1^2 + ... + \pa_D^2) - \frac{x_1}{a^2}\ga^1\ga^2\pa_2 -
\frac{x_1}{a^2}\ga^1\ga^3\pa_3 - ... - \frac{x_1}{a^2}\ga^1\ga^D\pa_D - m^2)\psi = 0.
\label{DEsquared}
\end{equation}
Obviously, we can seek solution in the form $\psi=\exp[-i\omega x_0 + ip_2x_2 + ... + ip_Dx_D]f(x_1)$
\begin{eqnarray}
&&\left[\omega^2 + \frac{(D-1)^2}{4a^2} - \frac{D-2}{a^2} x_1\pa_1 +
\frac{x_1^2\pa_1^2}{a^2}
- \frac{x_1^2}{a^2}(p_2^2 + ... + p_D^2) -i\frac{x_1}{a^2}\ga^1\ga^2p_2\right.\nonumber\\
&&\left.- i\frac{x_1}{a^2}\ga^1\ga^3p_3 - ... - i\frac{x_1}{a^2}\ga^1\ga^Dp_D -
m^2\right]f(x_1) = 0.
\end{eqnarray}
Further, $\gamma$ matrices are present only in the operator
\begin{eqnarray*}
-i\frac{x_1}{a^2}\ga^1\ga^2p_2 - i\frac{x_1}{a^2}\ga^1\ga^3p_3 - ... - i\frac{x_1}{a^2}\ga^1\ga^Dp_D.
\end{eqnarray*}
Since this is an Hermitian operator and its square $
{x_1^2}(p_2^2 + ... + p_D^2)/{a^4}$
is a unit matrix, it can be diagonalized and  its eigenvalues are equal to
$\sigma{x_1}\sqrt{p_2^2 + ... + p_D^2}/{a^2}$,
where $\sigma=\pm$. Making the change of variable $z=\sqrt{p_2^2+ ... + p_D^2}\,x_1$, we obtain
\begin{equation}
\left(\omega^2 + \frac{z^2}{a^2}(-1+\pa_z^2) + \frac{(D-1)^2}{4a^2}
 - \frac{(D-2)z}{a^2}\pa_z - \frac{\sigma z}{a^2} - m^2 \right)f(z) = 0.
\label{effectiveDE}
\end{equation}
The absence of any dependence on $p_2,...,p_D$ in this equation is remarkable because it
means that energy does not depend
on them, i.e., it is the same for any $p_2,..., p_D$. Eq.(\ref{effectiveDE}) has the form
of equation of a $(1 + 1)$-dimensional
problem and it is easy to find its spectrum $\omega = \pm \sqrt{\nu^2/a^2 + m^2}$, where
$\nu$ takes values in
$(0, +\infty)$. We would like to note that the effective dimensional reduction
$D + 1 \to 1 + 1$ for
fermions on hyperbolic spaces $H^D$ was observed in \cite{G2} by analyzing the
heat kernel of Dirac operator on these spaces, however, physical reasons for this
reduction remained unclear. Here, we see
that this reduction is connected with the effective (1 + 1)-dimensional form
(\ref{effectiveDE}) of the Dirac equation. From the mathematical
viewpoint, the reduction $D + 1 \to 1 + 1$ is related to the spherical and scale
symmetries of the
$H^D$ metric written in the Poincare patch. Indeed, the spherical symmetry of
the $x_2,...,x_D$ part of metric (\ref{HDmetric})
reduces the dependence of energy on $p_2,...,p_D$ to the dependence on the only
invariant $p^2=p_2^2+...+p_D^2$ and then the symmetry of
metric (\ref{HDmetric}) with respect to scale transformations $x_i \to \la x_i\,
(i=\overline{1,D})$ eliminates
any dependence on $p_2,...,p_D$ in Eq.(\ref{effectiveDE}) for eigenfunctions.

Actually there exists a physically even more transparent way to show the occurence
of the effective dimensional reduction on $H^D$.
According to \cite{APNY}, the effective dimensional reduction in the infrared region
takes place in a quantum problem only when the corresponding classical motion has a
bounded character with respect to the coordinates over which the dimensional
reduction occurs. For example, let us consider the dimensional reduction in constant
magnetic field. In this case, classically
a charged particle
moves on circular orbits in the plane perpendicular to the constant magnetic field.
Since the radius of its orbit is proportional
to energy, a charged particle can go to infinity only if it has infinite energy.
Therefore, a classical charged particle of
finite energy moves in a finite region of the plane perpendicular to the constant
magnetic field. This bounded
character of motion means that the system is effectively of finite size and translates
to the effective dimensional reduction by 2 units
in the infrared region in the quantum problem \cite{GMSh1}. Since we found the
effective universal dimensional
reduction $D + 1 \to 1 + 1$ for fermions on $H^D$ for any $D$, it is interesting
to consider the classical motion in hyperbolic spaces
and see whether it has also a bounded character with respect to the coordinates
over which the effective
dimensional reduction takes place.

The Lagrangian of free classical particle on $H^D$ reads
\beq
L = \frac{a^2}{x_1^2} \left( \dot{x}_{1}^2 + \dot{x}_{2}^2 + ... + \dot{x}_{D}^2 \right).
\eeq
One can solve classical equations of motion (actually, it is convenient to start with
the case of the Lobachevsky plane and the general case $D \ge 3$ can then be
deduced by using the spherical symmetry over the $x_2,...,x_D$ coordinates) and find
that the classical trajectories of motion in $H^D$ (geodesics)
have the form
\begin{eqnarray}
z(t) = \ln(x_1(t)) = \ln\frac{v}{A\cosh(vt + C_0)},\,\,\, x_2(t) =
\frac{vC_2}{A^2} \tanh(vt + C_0) + \tilde{C}_2,\,\dots,\,\, x_D(t)  =
\frac{C_D}{A^2} \tanh(vt + C_0) + \tilde{C}_D, \label{trajectories}
\end{eqnarray}
where $A^2=C_2^2+C_3^2+...+C_D^2$ and $v, C_0, C_2,\tilde{C}_2,...,$
$C_D,\tilde{C}_D$ are arbitrary constants (there are 2D of them and, for a
particular trajectory, they are fixed by initial conditions) and we introduced
also the coordinate $z=\ln x_1$ , which is in a certain sense more natural from
the viewpoint of metric (\ref{HDmetric}) because then the spatial interval has
the flat space form $dz^2$ with respect to motion in this coordinate. It is
easy to see from (\ref{trajectories}) that the motion with respect to $z$
coordinate has the same character as the usual flat space motion except a time
interval of order $1/v$. Indeed, it follows from (\ref{trajectories}) that the
classical particle moves like $z(t)=vt + C$ for $t \ll -\frac{1}{v}$. For $|t|
\le \frac{1}{v}$, its motion differs from the familiar inertial motion in flat
space. For $t \gg \frac{1}{v}$, the particle goes back to $-\infty$ (where it
started its motion) and its motion has the usual inertial character. On the
other hand, motion with respect to $x_2,...,x_D$ coordinates has a completely
different character. A particle is practically motionless for almost all period
of time except the time interval of order $1/v$ when it moves some finite
distance. One can calculate the contribution to the spatial interval connected
with motion in $x_i$ ($i=\overline{2,D}$) coordinates along geodesics
(\ref{trajectories}) and find that (unlike motion in $x_1$ coordinate) it is
always finite. For example, the spatial interval connected with motion in $x_2$
coordinate is equal to
$$
l_{x_2} = a \int_{-\infty}^{+\infty} \sqrt{\frac{\dot{x}_{2}^2}{x_1^2}} dt =
\frac{\pi a|C_2|}{A}.
$$
Therefore, motion in $x_2,...,x_D$ coordinates takes place in a finite region
of space defined by initial conditions. At this point similarity with the
classical motion of a charged particle in constant magnetic field is
transparent, where the particle moves in a bounded region of the plane
perpendicular to the magnetic field. For the $H^D$ case, the classical motion
has a bounded character in $D-1$ coordinates. Therefore, it is clear why there
is the effective dimensional reduction by 2 units in flat space in a constant
magnetic field and the universal reduction $D + 1 \to 1 + 1$ in hyperbolic
spaces $H^D$ in the corresponding quantum problems.


\begin{thebibliography}{99}

\bibitem{Fomin} P.I. Fomin, V.P. Gusynin, V.A. Miransky and Yu.A. Sitenko,
Rivista del Nuovo Cimento. 6 (1983) 1.
\bibitem{BCS} J. Bardeen, L.N. Cooper, J.R. Schrieffer, Phys. Rev. 108 (1957) 1175.
\bibitem{Alf}  M. Alford, K. Rajagopal, F. Wilczek, Phys. Lett. B 422 (1998) 247;
R. Rapp, T. Sch\"{a}fer, E.V. Shuryak, M. Velkovsky,  Phys. Rev. Lett. 81 (1998) 53.
\bibitem{Shankar}  R. Shankar,  Rev. Mod. Phys. 66 (1994) 129;
J. Polchinski, Proceedings of the 1992 TASI. Ed. by J. Harvey, J. Polchinski,
 Singapore: World Scientific, 1993, hep-th/9210046.
\bibitem{GMSh1}  V.P. Gusynin, V.A. Miransky , I.A. Shovkovy,   Phys. Rev. Lett. 73 (1994) 3499;
Phys. Lett. B 349 (1995) 477;
\\Phys. Rev. D52 (1995) 4747.
\bibitem{GMSh3} V.P. Gusynin, V.A. Miransky, I.A. Shovkovy, Phys. Rev. D 52 (1995) 4718;
Nucl. Phys. B 462 (1996) 249.
\bibitem{UPHZH}  V.P. Gusynin, Ukr. J. Phys. 45 (2000) 603.
\bibitem{G1}  E.V. Gorbar,  Phys. Lett. B 491 (2000) 305.
\bibitem{C1}  I. Sachs, A. Wipf, Phys. Lett. B 326 (1994) 105;
S. Kanemura, H.-Y. Sato,  Mod. Phys. Lett. A11 (1996) 785;
T. Inagaki,  Int. J. Mod. Phys. A 11 (1996) 4561;
E. Elizalde, S. Leseduarte, S.D. Odintsov, Yu.I. Shil'nov,  Phys. Rev. D 53 (1996) 1917;
T. Inagaki, K.-I. Ishikawa,  Phys. Rev. D 56 (1997) 5097.
\bibitem{Ina}  T. Inagaki, T. Muta, S.D. Odintsov,  Prog. Theor. Phys. Suppl. 127 (1997) 93.
\bibitem{G2}  E.V. Gorbar, Phys. Rev. D 61 (2000) 024013.
\bibitem{Odintsov}  D.M. Gitman, S.D. Odintsov, Yu.I. Shil'nov,  Phys. Rev. D 54 (1996) 2968.
\bibitem{Bulaev} D.V. Bulaev, V.A. Geyler, V.A. Margulis,
Physica. B 337 (2003) 180.
\bibitem{GusyninPRL2005}V.P.~Gusynin and S.~G.~Sharapov,  Phys. Rev. Lett. 95 (2005) 146801.
\bibitem{Vozmediano}  A. Cortijo, M.A.H. Vozmediano, Europhysics Letters. 77 (2007) 47002.
\bibitem{Ando} Y. Zheng and T. Ando, Phys. Rev. B 65 (2002) 245420;
 N.M.R. Peres, F. Guinea and A.H. Castro Neto, Phys. Rev. B 73 (2006) 125411.
\bibitem{Herbut} I.F. Herbut, Phys. Rev. Lett. 97 (2006) 146401;
Phys. Rev. B 75 (2007) 165411.
\bibitem{Comtet}  A. Comtet, P.J. Houston, J. Math. Phys. 26 (1985) 185.
\bibitem{Grad}  I.S. Gradsteyn, I.M. Ryzhik,  Tables of Integrals, Series,
and Products. - New York: Academic Press, 1965.
\bibitem{oscillations} S.G. Sharapov, V.P. Gusynin and H. Beck,  Phys. Rev. B 69 (2003) 075104.
\bibitem{Streda}P. Streda, J. Phys. C 15 (1982) L717.
\bibitem{Novoselov}  K.S. Novoselov,  et al., Nature. 438 (2005) 197;
Y. Zhang, et al., Nature. 438 (2005) 201.
\bibitem{Jellal} A. Jellal, preprint Arxiv:0709.4126,  2007.
\bibitem{Gross_Neveu} D. Gross and A. Neveu, Phys. Rev. D 10 (1974) 3235.
\bibitem{APNY}  D.J. O`Connor, C.R. Stephens, B.L. Hu, Ann. Phys. (NY). 190 (1989) 310.
\end{thebibliography}
\end{document}